\documentstyle[preprint,aps,psfig,tighten]{revtex}
\begin{document}
\draft
\title{Asymmetrically warped R=0 braneworlds}
\author{Sayan Kar \footnote{Electronic address: {\em sayan@cts.iitkgp.ernet.in}}
${}^{}$
}
\address{Department of Physics and Centre for
Theoretical Studies \\Indian Institute of Technology, Kharagpur 721 302, India}
\maketitle
\begin{abstract}
A new family of five dimensional, 
R=0 braneworlds with asymmetric warp factors  
is proposed. 
Beginning with the invariance of the Ricci scalar for the general class of 
asymmetrically warped spacetimes we, subsequently 
specialise to the $R=0$ case. Solutions  are 
obtained by choosing a particular relation (involving a
parameter $\nu$) between the warp factors. Symmetric
warping arises as a special case (particular value of $\nu$).
Over a range of values of $\nu$ the
energy conditions for the matter stress energy are found to hold good.
It turns out that 
the energy density and pressures required to support these spacetimes 
decay as the inverse square of the
fifth (extra) coordinate. The projection of this bulk stress--energy (for
symmetric warping) on the 3--brane yields an effective
cosmological constant. We conclude with brief comments on spacetimes
with constant Ricci scalar and the extension of our results
to diverse dimensions.    
\end{abstract}


\newpage

\maketitle
\vspace{.2in}
Ever since Kaluza and Klein {\cite{kk}} investigated five dimensional
General Relativity (GR) and derived its equivalence with four dimensional
GR coupled to electromagnetism, interest in extra dimensions have
drawn particle theorists and relativists towards the construction of
newer models {\cite{kkref}}. In the later part of the last century, 
extra dimensions
arose in the context of superstrings {\cite{string}} where their presence 
is unavoidable. Few years ago, the notion of nonfactorisable
spacetimes with an extra dimension, 
was initiated through the work of Randall and Sundrum {\cite{rs}}. 
Nonfactorisability of a line element in this context
essentially implies that the four dimensional part 
has a dependence on the fifth coordinate through an overall conformal factor. 
It was shown
that there exists an exact solution of the five dimensional 
Einstein equation with a 
negative cosmological constant, where the four dimensional part 
is Minkowski space with a conformal factor dependent on the fifth
coordinate (denoted henceforth as $\sigma$). This `warping' of the
four dimensional section results in a constant scaling of Minkowski
space--the scale factor taking on different values with different
choices of $\sigma$.
The Randall--Sundrum solution is given as :

\begin{equation}
ds^2 = e^{2f(\sigma)}\left [ -dt^2 + dx^2 +dy^2 + dz^2\right ] + r_c^2d\sigma^2
\end{equation}

where $f(\sigma) = -kr_c\vert \sigma \vert$ is the warp factor. 
The modulus in the form of
$f(\sigma)$ leads to the possibility of delta function sources in the
energy momentum tensor--this being attributed to the energy momentum due
to three--branes located at specific points along the extra dimension.
The motivation of this work was to arrive at a solution to the
heirarchy problem in particle physics. The RS solution sparked off
a flurry of papers on diverse topics associated with particle
physics {\cite{besan}}, cosmology {\cite{cosmo}}, black holes {\cite{black}}. 
The interest was largely due to the
claim that the presence of extra dimensions
could indeed be observed through experiments performed 
at the present-day energy scales achievable in 
high energy accelerators.

The line element mentioned above has one warp factor. In an attempt 
to generalise the above geometry, different warp factors for 
the spatial and temporal parts of the four dimensional section
were introduced in {\cite{csaki}}. Exact solutions in the presence
of a cosmological constant were reported and cosmological
consequences of such asymmetric warping were analysed. 

In this article we first recall that with 
such asymmetric  warping a vacuum solution
(without a cosmological constant) in five-dimensional GR is possible. 
This result, though straightforward, does not seem to find explicit mention
anywhere in the literature on braneworlds. It may be noted that setting the
cosmological constant to zero in the solution discussed in
{\cite{csaki}} will not yield the vacuum solution. We begin
with a brief analysis on the invariance of the Ricci scalar and
some qualitative remarks on the properties of the $R=0$ spacetimes
for asymmetric warping.
Subsequently, we move on towards constructing some specific 
asymmetrically warped
spacetimes with their Ricci scalar $R=0$ ($R_{ij}=G_{ij}
\neq 0$). A one parameter family of such solutions is obtained. The
status of the energy conditions for this class of solutions is
explored and some comments on them are outlined at the end.

We begin with the metric ansatz :

\begin{equation}
ds^2 = -e^{2f(\sigma)}dt^2 + e^{2g(\sigma)} \left (dx^2 + dy^2 +dz^2 \right )
+r_c^2 d\sigma^2
\end{equation}

where $f(\sigma)$ and $g(\sigma)$ are the warp--factors of the
time and space parts respectively, of the four dimensional section
of the full line element.

Let us write down the Einstein
tensor components
for the generic form of the metric given above : 
These (in the frame basis) are :

\begin{equation}
G_{tt} = - \frac{3g'' + 6 {g'}^2}{r_c^2}
\end{equation}
\begin{equation}
G_{xx}=G_{yy} = G_{zz} = \frac{2f'g'+3{g'}^2 +2 g'' + {f'}^2 + f''}{r_c^2}
\end{equation}
\begin{equation}
G_{\sigma\sigma} = \frac{3{g'}^2 + 3f'g'}{r_c^2}
\end{equation}

The first obvious question is : Is there a vacuum solution?
Notice that if we assume $f=g$ for all $\sigma$, the vacuum solution
of the equations $G_{ij} =0$ is trivial (f and g are constants). 
Therefore,
to have a nontrivial but symmetric warping
a negative cosmological constant is necessary. It is also
worth noting that the negativity of $\Lambda$ is a
necessity, in order to maintain consistency with the equation
$G_{\sigma\sigma} = 6 \frac{{g'}^2}{r_c^2}= -\Lambda$.

On the other hand, if we assume $f=-g$ for all $\sigma$, it is easy to
see that $G_{\sigma\sigma}=0$ straightaway. The other Einstein tensor
components when set to zero yield a single differential equation for the function
$g(\sigma)$ given as :

\begin{equation}
g'' + 2{g'}^2 =0
\end{equation}

Solving this equation we find 

\begin{equation}
g(\sigma) = \frac{1}{2}\ln (2\sigma + C)
\end{equation}

where $C$ is an arbitrary constant. For $C>0$ and $\sigma \geq 0$ we find
that the line element is a nonsingular vacuum solution.

The full line element therefore turns out to be :

\begin{equation}
ds^2 = -\frac{1}{2\sigma +C}dt^2 + (2\sigma + C)\left ( dx^2 + dy^2 + dz^2 
\right ) + r_c^2 d\sigma^2
\end{equation}

It is easy to check that $R=0$ and $R_{ij} =0$ for this solution. 
All Riemann tensor
components are finite as long as the limits on $C$ and $\sigma$ are
obeyed. It can be shown, not surprisingly that the above solution
is none other than the five dimensional Schwarzschild solution
where $\sigma$ is equivalent to the usual radial variable. 

Inclusion of a cosmological constant (i.e. solving the five
dimensional equations  $G_{ij} + \Lambda g_{ij} = 0$) leads to a solution
of the form :

\begin{equation}
ds^2 = -\frac{\sin^{2}2\alpha\sigma}{\cos 2\alpha\sigma} dt^2 + \cos2\alpha\sigma
\left [ dx^2 +dy^2 +dz^2\right ] + r_c^2 d\sigma^2
\end{equation}

where $\alpha = \sqrt{\frac{\Lambda r_c^2}{6}}$.

Using a coordinate transformation $\sigma ' = \cos 2\alpha\sigma$ we can 
rewrite
the above solution in the form :

\begin{equation}
ds^2 = -\frac{1-{\sigma'}^{4}}{{\sigma'}^2}dt^2 + {\sigma'}^2 \left [
dx^2 + dy^2 +dz^2 \right ]  + \frac{r_c^2}{\alpha^2} \frac{{\sigma'}^2}{1-
{\sigma'}^4} d{\sigma'}^2
\end{equation}

This above form has been derived in  
$\cite{csaki}$, $\cite{oda}$. It should be noted that the vacuum
solution {\em cannot} be arrived at from the above solution by choosing 
$\alpha =0$ (i.e. setting the cosmological constant to zero). However,
we mention that the above two solutions are {\em not} new and have been
discussed in the literature in the references cited above.

We now move on to constructing a new class of solutions 
for which $R_{ij}=G_{ij}\neq 0$ but $R=0$. This is inspired, in a
sense, by the fact that in spherically symmetric, static General Relativity, 
the full family of $R=0$ solutions includes Schwarzschild (vacuum),
Reissner--Nordstrom (non--vacuum, electric field present) and 
a host of other solutions recently discussed by this author, along with
others {\cite{R=0}}. We make an attempt to find a class of
nonfactorisable spacetimes in five dimensions, subject to  
the $R=0$ constraint. 
From the expression for the Einstein tensors quoted above it is easy
to write down the Ricci scalar, which, when set equal to zero, yields the
equation : 

\begin{equation}
3g'' + 6{g'}^2 + 3f'g'+{f'}^2 + f'' =0
\end{equation}

The above equation cannot be solved without imposing a 
specific relation between $f$ and $g$. However, we can make some
general comments about the nature of the functions $f$ and $g$ by
inspecting the above constraint. Firstly, if at some value, say
$\sigma_0$, $f$ and $g$ both have an extremum (which implies
$f'(\sigma_0)=0, g'(\sigma_0)=0$) then $f''(\sigma_0) = -3g''(\sigma_0)$.
Thus, at $\sigma_0$ if $f$ has a maximum then $g$ will have a minimum
and vice versa. Secondly, if $f$ has an inflexion point at some
$\sigma_0$ (i.e. $f'$ and $f''$ are both zero), then $g''$ is either 
negative ($g' (\sigma_0)\neq =0$ or $g$ has an inflexion point.   

Furthermore, one may investigate the symmetries of the Ricci
scalar (which is equal to -2 times the L.H.S. of the Eqn. (11) above)
for the general class of asymmetrically warped spacetimes. 
Let us consider the following linear transformation in f--g space: 
$f \rightarrow p f + q g$, $g \rightarrow r f + s g$, where
$p,q,r,s$ are real numbers. It turns out that the Ricci scalar
remains {\em invariant} only under the following values of (p,q,r,s):
($-\frac{1}{2}, \frac{3}{2}, \frac{1}{2}, \frac{1}{2}$) and 
($1,0,0,1$). The first of these is nontrivial whereas the second one
is just the identity. Viewing the transformation via a matrix relation it
is easy to see that the symmetry group is isomorphic to $Z_2$. 
Two succesive applications of the first transformation yields the
identity. The determinant of the matrix representing the first transformation
is $-1$. Therefore, it is an
improper transformation in f--g space and different from rotations. 
It is possible to utilise the transformation to
generate new solutions with the same value of the Ricci scalar. $R=0$
is just one case which we shall discuss below. As an example, one may
start out with the vacuum solution discussed above 
and generate a solution with
nontrivial matter stress energy but with the same value of the 
Ricci scalar. This is reminiscent of a duality proposed a few years
ago by Dadhich {\cite{dadhich}} though there it wasn't very clear how to 
implement the
transformation at the level of the metric.   

Let us now assume the simple relation
$f=\nu g$ ($\nu$ a finite, non--zero constant). This, certainly is
a {\em choice} made in order to solve the $R=0$ constraint. It is
qualitatively similar, for instance, to the choice of the Schwarzschild gauge
$g_{00} = -[g_{11}]^{-1}$, for which the $R=0$ constraint yields
both the Schwarzschild and the Reissner--Nordstrom solutions.

The $R=0$ condition for our case becomes :

\begin{equation}
g'' + \eta {g'}^2 = 0
\end{equation}

with $\eta = \frac{\nu^2+3\nu+6}{\nu+3}$ (with $\nu \neq -3$). 
For $\nu=-3$ the above constraint yields $g=f=constant$, which is
a trivial solution.
The general solution to the
above constraint turns out to be :

\begin{equation}
g(\sigma)=\frac{1}{\eta}\ln (\eta \sigma +C)
\end{equation}

Therefore, using $f=\nu g$ we get

\begin{equation}
f(\sigma) = \frac{\nu}{\eta} \ln (\eta\sigma + C)
\end{equation}

The line element turns out to be :

\begin{equation}
ds^2 = -\left (\eta\sigma + C\right )^{\frac{2\nu}{\eta}} dt^2 + \left (\eta\sigma +
C\right )^{\frac{2}{\eta}} \left [dx^2+dy^2+dz^2\right ] + r_c^2d\sigma^2
\end{equation}

The matter stress energy which is required to generate this solution 
is obtained
from the Einstein tensors for this line element, via the five dimensional
Einstein equations. Defining $\rho$, $p$ and $p_\sigma$ as the diagonal
components of the five dimensional energy--momemtum tensor, we have

\begin{eqnarray}
8\pi {G_5} \rho = G_{tt} = 
\frac{3}{r_c^2} \frac{\nu(\nu+1)}{(\nu+3)} \frac{1}{(\eta\sigma + C)^2} \\
8\pi {G_5}p = G_{ii} = -\frac{3}{r_c^2}\frac{(\nu+1)}{(\nu+3)} \frac{1}{
(\eta\sigma + C)^2}\\
8\pi {G_5}p_\sigma = G_{\sigma\sigma} = 
\frac{3}{r_c^2} \frac{(\nu+1)}{(\eta\sigma + C)^2} \\
\end{eqnarray}

where $G_5$ is the five dimensional gravitational constant.

The equation of state for the matter in the bulk would therefore be:

\begin{equation}
p = -\frac{1}{\nu} \rho \hspace{.2in} ; \hspace{.2in} p_\sigma = \frac{\nu+3}{\nu}\rho
\end{equation}

The pressures in the three spatial and the extra $\sigma$ direction
have equations of state which resemble the well--known ones of the form 
$p=\gamma \rho$. Here we have different values of $\gamma$ 
for the pressures along the spatial (x,y,z) and $\sigma$ directions.
Note also that the it is possible for the pressures along the spatial and
$\sigma$ directions to be of opposite sign. The bulk stress energy is
therefore not a perfect fluid though the projection on the brane has this
character.   
 
It is worth noting that for $\nu=1$, the asymmetry in the spatial
and temporal warp factors is no longer there and we have a Randall--Sundrum
type model with a line element given as :

\begin{equation}
ds^2 = (\frac{5}{2}\sigma + C)^{\frac{4}{5}}\left (-dt^2 + dx^2 +
dy^2 +dz^2 \right ) + r_c^2 d\sigma^2
\end{equation} 

We may also rewrite the above line element in a conformally flat form
by using a coordinate transformation $\sigma' = \frac{2}{3}r_c (\frac{5}{2}
\sigma + C)^{\frac{3}{5}}$. The line element therefore becomes:

\begin{equation}
ds^2 = \left (\frac{3\sigma'}{2r_c}\right )^{\frac{4}{3}} \left ( -dt^2 +
dx^2 +dy^2 +dz^2 + d{\sigma'}^2 \right )
\end{equation}

However, the matter source here is different. The bulk 
negative cosmological constant is no longer present.
The equation of state with this choice of $\nu$ becomes : 
$p=-\rho, p_\sigma = 4\rho$.
In general, for all $\nu$, the matter stress--energy 
varies with $\sigma$ as $\frac{1}{\sigma^2}$, thereby,
progressively decreasing as we move further and further away along $\sigma$. 
For $\nu =0$ we find $f=0$ but $g= \frac{1}{2}\ln (\eta \sigma + C)$
This spacetime has the strange feature that it possesses zero energy
density but non--zero partial presssures. 
For $\nu =-1$ we obtain the vacuum solution mentioned earlier in this
article.

Let us now concentrate on the various energy condition inequalties
that exist in the literature {\cite{wald}}, {\cite{hawk}}
and examine their status in the
context of the one--parameter family of solutions obtained so far.
A word about energy conditions will not be inappropriate here.
These conditions are constraints on matter stress energy which
are dictated by physical requirements in the classical world.
For instance, in one of the conditions it is necessary that
the energy density $\rho \ge 0$--which is obvious as long as
we describe matter classically (as opposed to quantum
expectation values of stress energy). It is widely
accepted that if the energy conditions are satisfied 
then we can say that the matter stress--energy is {\em reasonable}
even though particular field theoretic models for them may not
be available.

The Weak Energy Condition (WEC) for a diagonal energy--momentum tensor
of the kind mentioned earlier reduces to the following inequalties:

\begin{equation}
(i)\hspace{.2in} \rho \ge 0 \hspace{.2in} (ii) \hspace{.2in} \rho+ p \ge 0
\hspace{.2in} (iii) \hspace{.2in} \rho + p_{\sigma} \ge 0
\end{equation}

A weaker set of inequalties containing only (ii) and (iii) of the set for
WEC comprises the Null Energy Condition (NEC). 

For the solution under consideration we therefore require (for WEC):

\begin{equation}
(i)\hspace{.2in} \rho \ge 0 \hspace{.2in} (ii) \hspace{.2in} (1-\frac{1}{\nu})\rho \ge 0  \hspace{.2in}
(iii)\hspace{.2in} \frac{(2\nu+3)}{\nu}\rho \ge 0
\end{equation}

The NEC will therefore consist of only (ii) and (iii) of the above.

Using the Einstein equations one may replace $\rho$ by $G_{00}$ and
proceed towards finding the domains of $\nu$ for which these
inequalities may be satisfied.

From the expression for $\rho$ mentioned earlier it is easy to see that
$\rho \ge 0$ if $-3<\nu\le -1$ and $\nu \ge 0$. However, (ii)
will require $\nu \ge 1$ or $\nu < 0$. Similarly, (iii) 
implies $ \nu > 0$ or $ \nu \le  -\frac{3}{2}$. Combining the three
we find that {\em WEC can be satisfied only if $-3 < \nu \le -\frac{3}{2}$
or $\nu \ge 1$}. We can also check that if $\rho$ is negative it is
not possible to satisfy the other inequalities in order to conserve only
the NEC.  

A couple of other energy conditions also exist in the literature and
are widely used. These are the Strong Energy Condition (SEC) and the
Dominant Energy Condition (DEC).  The relevant inequalities for DEC and
SEC are :

\begin{eqnarray}
SEC : \hspace{.2in} (i)\hspace{.2in} \rho + p,p_\sigma \ge 0 \hspace{.2in}
(ii)\hspace{.2in} \rho + 3p + p_\sigma \ge 0  \\ 
DEC : (i) \hspace{.2in} \rho \ge 0 \hspace{.2in} \& \hspace{.2in} -\rho \le p,p_\sigma \le \rho
\end{eqnarray}

Using the equation of state for the $R=0$ solutions we find that :

\begin{eqnarray}
SEC \Rightarrow  (i)\hspace{.2in} (1-\frac{1}{\nu})\rho \ge 0 \hspace{.2in} 
(ii) \hspace{.2in} \frac{2\nu+3}{\nu} \rho \ge 0 (iii) \hspace{.2in} 6 \rho \ge 0\\
DEC \Rightarrow  (i) \hspace{.2in} \rho \ge 0 \hspace{.2in} (ii) \hspace{.2in}
-2 \le -\frac{1}{\nu} \le 0 \hspace{.2in} (iii)\hspace{.2in}  -2 \le \frac{3}{\nu} \le 0 
\end{eqnarray}

It is clear that DEC cannot be satisfied for any domain of $\nu$.
whereas if WEC holds then SEC also holds.
{\sf In summary WEC and SEC can hold over the domain 
$-3 <\nu\le -\frac{3}{2}$ and $\nu \ge 1$.}
If $\nu=-1$,then all energy conditions hold because this
corresponds to vacuum. For $\nu=-4$ it is possible to have
isotropic pressures $p.p_\sigma = \frac{\rho}{4}$. However $\rho$
is negative for this value of $\nu$.

Several other comments on these new $R=0$ solutions are in order.
We discuss them pointwise below. 

(a) The asymmetry in the warp factors implies that an arbitrary constant 
$\sigma$ section will introduce 
different scale factors for the time and space parts of the
four dimensional line element. This will obviously lead to
violation of global Lorentz invariance since each constant $\sigma$
section will have a different scaling for the time and space
coordinates. Hence the velocity of light will change as we
move from one slice to another, implying that the presence
of the extra dimension acts as a medium with varying refractive
index.
To avoid this perhaps undesirable feature, we have to confine
ourselves to a $\sigma = constant$ slice for which the
space and time scalings are the same and the metric on the
slice is just Minkowski. 
This can be achieved if we choose $\sigma = \sigma_0 = \frac{1 -C}{\eta}$
By this choice the metric on the 3--brane section becomes the
Minkowski metric and we can claim that this is our usual four--dimensional
world where we live.  

(b) Following Randall--Sundrum it is also possible to introduce
$\delta$--function sources representing the brane matter localised
on the domain walls. This could be done by replacing $\sigma$ by
$\vert \sigma \vert$. 

(c) The energy momentum on the
3--brane section for the $\nu =1$ line element, 
has the property that $\rho = -p$ where $\rho$
is the energy density and $p$ are the pressures along the spatial
directions. This is the equation of state for stiff matter. In other
words, since the matter stress energy depends only on the fifth coordinate,
the projection on the 3--brane yields an effective negative cosmological
constant in the brane world. Such a cosmological constant could be
obtained from a massless scalar field living on the 3--brane. 

(d) It is easy to check that if we assume $f(\sigma) = k_1\sigma^n$ and
$g(\sigma) = k_2\sigma^n$ for any $n \neq 0 $ the $R=0$ condition cannot
be satisfied.  

(e) It is possible to obtain $R=\Lambda$ (where $\Lambda$ is a constant) spacetimes by replacing the
Eqn (12) above with the condition $g''+\eta {g'}^2 = K$ ($K =-\frac{\Lambda
r_c^2}{2}$) 

 . For 
positive and negative cosmological constants ($\Lambda$) the warp factors
turn out to be :

{\bf $\Lambda < 0, K>0 $ : }

\begin{equation}
e^{2f(\sigma)} = \left (\cosh ( a\eta\sigma ) \right )^{\frac{2\nu}{\eta}}
\hspace{.1in} ; \hspace{.1in} e^{2g(\sigma)} = \left (\cosh (a\eta
\sigma \right )^{\frac{2}{\eta}}
\end{equation}

{\bf $\Lambda >0, K<0$ :}

\begin{equation}
e^{2f(\sigma)} = \left (\cos a\eta (\sigma_0 -\sigma ) \right )^{\frac{2\nu}{\eta}}
\hspace{.1in} ; \hspace{.1in} e^{2g(\sigma)} = \left (\cos a\eta
(\sigma_0-\sigma ) \right )^{\frac{2}{\eta}}
\end{equation}

where $ a = \sqrt{{K}/\eta}$ for the first solution and 
$a=\sqrt{\vert K \vert /\eta}$ for the second one.

These are constant Ricci scalar line elements and have asymmetric warping
except for the case when $\nu=1$. The matter stress energy required to
generate these spacetimes would be more complicated and we shall not
delve any further with these solutions. 

(f) It is easy to generalise the above solutions to higher or
lower dimensions. For example one may look at the line element in
D dimensions :

\begin{equation}
ds^{2} = e^{2f}dt^2 + e^{2g}\left [dx_1^2+ dx_2^2 + ... dx_{D-2}^2 
\right ] +r_c^2 d\sigma^2
\end{equation}

The form of the $R=0, R_{ij}=G_{ij}\neq 0$ solution remains the same 
(i.e. $f(\sigma) = \nu g(\sigma) = \frac{\nu}{\eta} \ln (\eta\sigma +C)$) 
for this D dimensional
geometry. However, the quantity $\eta$ turns out to be :

\begin{equation}
\eta = \frac{\left ( 2 \nu^2 + 2(D-2)\nu + (D-1)(D-2)\right )}{2\left ( \nu + (D-2)\right )}
\end{equation}

For $\nu =1$ (symmetric warping) we find that $\eta =\frac{D}{2}$ and the stress--energy is
given as :

\begin{equation}
p = -\rho = -\frac{p_{\sigma}}{D-1} = -\frac{D-2}{2r_c^2}\frac{1}{(\eta \sigma
+C)^2}
\end{equation}

On the other hand for $\nu =-1$ we have 

\begin{equation} 
p=\rho = -\frac{p_{\sigma}}{D-3}= -\frac{(D-2)(D-5)}{2r_c^2 (D-3)}\frac{1}{
(\eta\sigma + C)^2}
\end{equation}

We notice that a vacuum solution with asymmetric warping exists only in $D=5$.
$D=3$ is another special case which needs to be treated separately. Note that
for $D=3$ $\nu=-1$ is not an useful choice because it yields constant 
values for $g$ and hence $f$. With $D=4$ and $\nu =1$ the metric coefficients turn out
to be linear in $\sigma$ (i.e. $e^{2g} = 2\sigma + C$).  
The violation/conservation of energy conditions
can be worked out without much difficulty--we do not indulge in deriving
them explicitly here. 

We now conclude with a summary of the results obtained.
It has been shown that there exists a one--parameter family of
nonfactorisable spacetimes with asymmetric warp factors in general
and symmetric ones in particular, for which the energy conditions
of GR can be satisfied. These solutions present an alternative to
the Randall--Sundrum type geometries. The absence of a cosmological
constant in the bulk is however replaced by the presence of a bulk
stress energy. The projection of this stress energy for the symmetrically
warped case shows that we have an effective cosmological constant on the
brane. The stress energy, as we move along the extra dimensions decays
quadratically with distance and goes to zero as we approach infinity.
In a sense the stress energy is like the one we have in the standard
model in cosmology where the inverse--square temporal dependence appears
for scale factors of the form $t^{\beta}$. Therefore, the five
dimensional warped geometry behaves like a `cosmological model' where
the extra dimension is like the `time' coordinate and the warp factor
is reminiscent of the `scale factor'. This visualisation is possible
if we realise that the homogeneous isotropic cosmological line element
is also, in a sense `warped' with the scale factor as the warp factor
and the time coordinate, an extra dimension.

Besides the geometries with $R=0$ obtained in the above, we have also found a 
symmetry of the Ricci scalar for the whole class of asymmetrically warped
spacetimes. This symmetry, which is essentially isomorphic to  $Z_2$, 
enables us to
generate new spacetimes with the same Ricci scalar (but
different matter stress energy) from a given one. It would be worthwhile
to explore the consequences of this symmetry in diverse contexts in future.  
 
Apart from obtaining these solutions and investigating the status of
the energy conditions  we have also examined the possibility of
a spacetime with constant Ricci scalar. Finally, we comment on
the extension of these solutions for $D$ dimensions. It turns out
that the form of the solution as well the stress energy remains the same  
, except for the dependence of the quantity $\eta$ on $D$ and hence
its different values for different $D$. Additionally, we obtain the
curious result that $D=5$ is an unique case for which a vacuum
spacetime with asymmetric warp factors exists. We mention, however, that
we have not discussed a viable model for the stress--energy tensor 
with matter fields--though it should be possible to do so using
scalar or vector fields. A further line of generalisation could be
the introduction of nontrivial curvature in the 3--brane section-- i.e.
by replacing the Minkowski metric by a metric with nonzero curvature. 

We leave behind several unanswered questions. Is this model useful
for the solution of the heirarchy problem? 
What are the corrections to Newtonian gravity in this model? How would the
solutions change if one assumes a metric with a nontrivial
curvature on the 3--brane section?  We have, in
this paper, obtained some solutions. It will be necessary to
address the abovestated questions in order to be sure that the
model is useful. However, even without its use in solving the problems
mentioned, the geometries remain as possible exact solutions with the
simple geometric property that the Ricci scalar is zero. In addition,
the invariance of the Ricci scalar obtained above is also a 
fact worth noting.

\end{document}